\documentclass[prd,superscriptaddress,amsmath,notitlepage,twocolumn]{revtex4-1}
\usepackage{graphicx}
\usepackage{float}
\usepackage{epstopdf,cancel}
\usepackage{epsf,latexsym,bbm,euscript}
\usepackage{amssymb,amsmath}
\usepackage{mathtools} %used in this file for the command \coloneqq -> :=
\usepackage{times,graphics}
\usepackage{soul,xcolor}
\usepackage{mathtools}

\def\6{{\langle}}
\def\9{{\rangle}}
\newcommand{\defeq}{\vcentcolon=}

\newcommand{\be}{\begin{equation}}
\newcommand{\ee}{\end{equation}}
\newcommand{\ba}{\begin{eqnarray}}
\newcommand{\ea}{\end{eqnarray}}

\def\half{{\tfrac{1}{2}}}

\def\pad{{\partial}}

\def\sg{\textsl{g}}

 \def\eF{\EuScript{F}}
 \def\eL{\EuScript{L}}
\def\cO{\mathcal{O}}

 \def\rE{\mathrm{E}}

\def\maf{\mathfrak{f}}

\def\vS{v_\mathrm{S}}

\usepackage{url,hyperref}
\hypersetup{colorlinks,linkcolor={blue!55!black},citecolor={red!45!black},urlcolor={blue!45!black},breaklinks=true}

\begin{document}

\title{Universal properties of the near-horizon geometry}

\author{Sebastian Murk}
\affiliation{Department of Physics and Astronomy, Macquarie University, Sydney, NSW 2109, Australia}
\affiliation{Sydney Quantum Academy, Sydney, NSW 2006, Australia}
%\email{sebastian.murk@mq.edu.au}

\author{Daniel R. Terno}
\affiliation{Department of Physics and Astronomy, Macquarie University, Sydney, NSW 2109, Australia}
% \email{daniel.terno@mq.edu.au}

\begin{abstract} 
	We derive universal properties of the near-horizon geometry of spherically symmetric black holes that follow from the observability of a regular apparent horizon. Only two types of solutions are admissible. After reviewing their properties we show that only a special form of the solutions of the second type is consistent. We describe how these results extend to modified theories of gravity, including Einstein\textendash{}Cartan theories. Then we describe the unique black hole formation scenario that necessarily involves both types of solutions. The generalized surface gravity is infinite at the apparent horizon. This feature and comparison of the required energy and timescales with the known semiclassical results suggest that the observed astrophysical black holes are horizonless ultra-compact objects, and the presence of a horizon is associated with currently unknown physics. 
\end{abstract}

\maketitle

\section{Introduction}
Models of astrophysical black holes describe them  as ultra-compact objects (UCOs) with or without a horizon \cite{cp:na17}. Both types of models are consistent with current data \cite{gw-12:rev,eht-1:19} and combine theoretically appealing properties with arguably undesirable features. 

Spherical symmetry considerably simplifies the analysis. Nonetheless, %even within the semiclassical framework
definite results are obtained only if the spacetimes under consideration are static, either exactly or asymptotically, and/or matter follows a prescribed evolution \cite{b:book,isr,he:book,fn:book,p:book}. Numerical studies must assume the matter content and the equations of state \cite{num:book}. As a result, despite spectacular successes in modeling the behavior of UCOs, the question of whether or not  they {actually}  have horizons  is still open \cite{cp:na17,bh-map,curiel}.

There is no unanimously agreed upon definition of a black hole \cite{curiel}, but strong gravity that locally prevents light from escaping is its most common characteristic. A physical black hole (PBH) \cite{f:14} has a trapped region, i.e.\ it contains a spacetime domain where ingoing and outgoing future-directed null geodesics originating from a two-dimensional spacelike surface with spherical topology have negative expansion \cite{he:book,fn:book,faraoni:b}. Its evolving outer boundary is the apparent horizon. In spherical symmetry it is unambiguously defined in all foliations that respect this symmetry.
Hence from the observational point of view \cite{visser,bmmt:19} a UCO is a PBH only if the apparent horizon has formed prior to emission of the signals that are detected by a distant observer (Bob). Nevertheless, this conceptual difference between a finite formation time of the trapped region and an asymptotically exponential approach to the event horizon of classical black holes translates into potentially observable signatures that arise from the different near-horizon properties that we present below.

Universal scaling laws were found in the formation of spherically symmetric black holes by looking at the late time limits of various types of collapsing classical matter \cite{universal-1}. All static Killing horizons \cite{he:book,fn:book,faraoni:b} lead to universal properties of the near-horizon optical geometry \cite{gw:09}. Our goal is to obtain the universal properties of PBHs that distinguish them from the alternative UCOs. These differences in geometry are expected to be resolved in future observations \cite{cp:na17,bh-map}.

Working in the framework of semiclassical gravity \cite{pp:09,bmt-1,hv:book}, we use classical notions (horizons, trajectories, etc.) and describe dynamics via the Einstein equations $G_{\mu\nu}=T_{\mu\nu}$ or modifications thereof. Our only assumption is that the apparent horizon that has been formed at some finite time of Bob is a regular surface in the sense that the curvature invariants there are finite. Within any given theory, we do not assume any specific matter content nor a specific quantum state $\omega$ that produces the expectation values of the energy-momentum tensor (EMT) $T_{\mu\nu} = \6 \hat{T}_{\mu\nu} \9_\omega$. Note that this EMT describes the total matter content --- both the original collapsing matter and the produced excitations. We do not assume the presence of Hawking-like radiation, event horizon, or singularity.

\section{Admissible solutions}
We establish our results by constructing finite invariants $\mathrm{T}\defeq T^\mu_{~\mu}$  and $\mathfrak{T}\defeq T^{\mu\nu}T_{\mu\nu}$ from divergent quantities \cite{bmmt:19}. This behavior manifests itself in Schwarzschild coordinates, where a general spherically symmetric metric is given by
\be
ds^2=-e^{2h(t,r)}f(t,r)dt^2+f(t,r)^{-1}dr^2+r^2d\Omega. \label{sgenm}
\ee
These coordinates provide geometrically preferred foliations with respect to Kodama time, a natural divergence-free preferred vector field \cite{faraoni:b,k:80,av:10}. Using the advanced null coordinate $v$ the metric is written as
\be
  ds^2=-e^{2h_+}\left(1-\frac{C_+}{r}\right)dv^2+2e^{h_+}dvdr +r^2d\Omega . \label{lfv}
\ee
The Misner\textendash{}Sharp (MS) mass \cite{faraoni:b,ms} $C(t,r)$ is invariantly defined via
\be
	f(t,r) \defeq 1-C/r \defeq \pad_\mu r\pad^\mu r, \label{defMS}
\ee
and thus $C(t,r)\equiv C_+\big(v(t,r),r\big)$. The functions $h(t,r)$ and $h_+(v,r)$ play the role of integrating factors in coordinate transformations, such as
   \be
dt=e^{-h}(e^{h_+}dv- f^{-1}dr). \label{intf}
\ee
The apparent horizon is located at the Schwarzschild radius $r_\sg(t)\equiv r_+(v)$ that is the largest root of $f(t,r)=0$ \cite{faraoni:b,aphor}.

It is convenient to introduce
\be
 \tau_t\defeq e^{-2h} T_{tt}, \qquad \tau^r\defeq T^{rr}, \qquad \tau_t^{~r} \defeq e^{-h}T_t^{~r}.
\ee

Unless stated otherwise the dynamics is governed by the standard Einstein equations of general relativity (GR). The three Einstein equations for $G_{tt}$, $G_t^{~r}$, and $G^{rr}$ are
\begin{align}
	\partial_r C &= 8 \pi r^2 \tau_t / f , \label{gtt} \\
	\partial_t C &= 8 \pi r^2 e^h \tau_t^r , \label{gtr} \\
	\partial_r h &= 4 \pi r \left( \tau_t + \tau^r \right) / f^2 . \label{grr}
\end{align}

The two invariants satisfy $\mathrm{T}\equiv - {R}/8\pi$ and $\mathfrak{T}\equiv R^{\mu\nu}R_{\mu\nu}/64\pi^2$, where $R_{\mu\nu}$ and $R$ are the Ricci tensor and Ricci scalar, respectively. Since the term $T^\theta_{\,\theta}\equiv T^\varphi_{\,\varphi}$ is finite in GR  \cite{bmmt:19,t:20}, regularity of the apparent horizon requires that the scalars
\be
     \mathrm{T}=(\tau^r-\tau_t)/f, \qquad \mathfrak{T}= \big((\tau^r)^2+(\tau_t)^2-2(\tau^r_{t})^2\big)/f^2, \label{twoscal}
\ee
are finite at $r=r_\sg$. It was shown that only two classes of dynamic solutions (with the leading terms in the functions $\tau_a$, $a \in \lbrace t, tr, r \rbrace$ scaling as $f^k$, $k=0, 1$) satisfy the regularity conditions \cite{t:20}. After reviewing the properties of these solutions we demonstrate that for $k=1$ only  a specific solution is consistent, provide the explicit form of the coordinate transformation \eqref{intf} near $r_\sg$, demonstrate divergence of the generalized surface gravity, and discuss the implications for PBH formation and modified theories of gravity.

\subsection{Generic solution}
The solutions with $k=0$ allow $\tau_t \to \tau^r \to \mp \Upsilon^2$, $\tau_t^r\to\mp\Upsilon^2$ for some $\Upsilon(t)>0$, but only
\begin{align}
	&\tau_t\approx\tau^r=-\Upsilon^2+\cO(\sqrt{x}), \label{eq:taut+r} \\
	&  \tau^r_{t}=-\Upsilon^2+\cO(\sqrt{x}),   \label{eq:taus}
\end{align}
where $x\defeq r-r_\sg$, yield valid PBH solutions: taking $\tau_t \to \tau^r \to + \Upsilon^2$ results in complex-valued solutions of Eqs.~\eqref{gtt}--\eqref{grr} (see Ref.\ \cite{bmmt:19}, Appendix C). The negative sign of $\tau_t$ and $\tau^r$ leads to the violation of the null energy condition (NEC) \cite{he:book,p:book,mmv:17,ks:20} in the vicinity of the apparent horizon. A future-directed outward (inward) pointing radial null vector $k^\mu$ satisfies $T_{\mu\nu}k^\mu k^\nu<0$ for the contracting (expanding) Schwarzschild radius $r_\sg$ \cite{bmmt:19}. Accreting Vaidya black hole solutions in $(v,r)$ coordinates satisfy
the NEC and therefore do not describe PBHs (see Appendix \ref{aA} for details).

The metric functions that solve Eq.~\eqref{gtt} and Eq.~\eqref{grr} are
\begin{align}
	\begin{aligned}
		C &= r_\sg-4\sqrt{\pi} r_\sg^{3/2}\Upsilon\sqrt{x}+\cO(x) , \\ 
		h &= -\frac{1}{2}\ln{\frac{x}{\xi}}+\cO(\sqrt{x}) ,  \label{k0met}
	\end{aligned}
\end{align}
where $\xi(t)$ is determined by the choice of time variable and the higher-order terms depend on the higher-order terms in the EMT expansion \cite{bmt:19}. Eq.~\eqref{gtr} must then hold identically. Both sides contain terms that diverge as $1/\sqrt{x}$, and their identification results in the consistency condition
\be
	r'_\sg/\sqrt{\xi}=\mp4\sqrt{\pi r_\sg}\,\Upsilon. \label{rpr1}
\ee

Useful information can be obtained by working with retarded and advanced null coordinates. If $r'_\sg>0$, it is convenient to use the retarded null coordinate $u$. For $r'_\sg<0$, the advanced null coordinate $v$ is particularly useful \cite{bmmt:19},
\begin{align}
    C_+(v,r)&=r_+(v)+\sum_{i\geqslant 1} w_i(v)(r-r_+)^i,  \label{cv1} \\
    h_+(v,r)&=\sum_{i\geqslant 1} \chi_i(v)(r-r_+)^i, \label{hv1}
 \end{align}
for some functions $w_i(v)$, $\chi_i(v)$, where $w_1 \leqslant 1$ due to the definition of $r_+$. This is the general form of the metric functions in $(v,r)$ coordinates that ensures finite curvature scalars at the apparent horizon \cite{t:20}. In this case the components of the EMT in $(v,r)$ and $(t,r)$ coordinates are related by
\begin{align}
&\theta_v\defeq e^{-2h_+}\Theta_{vv}=\tau_t,  \label{thev}\\
&\theta_{vr}\defeq e^{-h_+}\Theta_{vr} = \left( \tau_t^r-\tau_t \right) / f , \label{thevr}\\
&\theta_r\defeq \Theta_{rr} = \left( \tau^r+\tau_t-2\tau^r_t \right) / f^2,   \label{ther}
\end{align}
where $\Theta_{\mu\nu}$ denotes the EMT components in $(v,r)$ coordinates.

A static observer finds that the energy density $\rho=T_{\mu\nu}u^\mu u^\nu=-T^t_{~t}$, pressure $p=T_{\mu\nu}n^\mu n^\nu=T^r_{~r}$, and flux $\phi \defeq  T_{\mu\nu}u^{\mu}n^\nu$, where $u^\mu$ is the four-velocity and $n^\mu$ is the outward-pointing radial spacelike vector, diverge at the apparent horizon. The experience of a radially-infalling observer Alice moving on the trajectory $x^\mu_\mathrm{A}(\tau)=(t_\mathrm{A}, r_\mathrm{A},0,0)$ is different, and also differs from the infall into a classical eternal black hole.

First, horizon crossing happens not only at some finite proper time $\tau_0$, but also at a finite time $t_0(\tau_0)$, $r_\sg\big(t_0(\tau_0)\big)=r_\mathrm{A}(\tau_0)$ according to the clock of a distant Bob. This is particularly easy to see for ingoing null geodesics, where
\be
 \frac{dt}{dr}= - \frac{e^{-h(t,r)}}{f(t,r)} \to \pm\frac{1}{r_\sg'}, \label{dv=0}
\ee
at $r=r_\sg$, the rhs is obtained by using Eqs.~\eqref{k0met} and \eqref{rpr1} \cite{t:20}, and the upper sign corresponds to evaporation. We use this result in the estimates of the formation time in Sec.~\ref{disc} and in showing that the usual generalizations of the surface gravity to nonstationary spacetimes fail for PBHs (Sec.~\ref{surfg}).

For an evaporating black hole ($r_\sg'<0$), energy density, pressure, and flux in Alice's frame are finite. However, upon crossing the apparent horizon of an accreting PBH, Alice encounters a firewall,
\be
 \rho_{\mathrm{A}}=T_{\mu\nu}u^\mu_\mathrm{A}u^\nu_\mathrm{A}=-\frac{  \dot{r}_\mathrm{A}^2}{4\pi r_\sg X} +\cO(1/\sqrt{X}), \label{fire}
\ee
where $X \defeq r_A(\tau)-r_\sg\big(t_\mathrm{A}(\tau)\big)$ \cite{t:19}.

Violations of the NEC are bounded by quantum energy inequalities \cite{ks:20,few:17}. Outside the singularities the lower bounds were shown to exist for the energy density   $\6 \hat{T}_{\mu\nu}\9_\omega u^\mu u^\nu$ and its smeared averages. These are known to be state-independent for free fields. For spacetimes of small curvature, explicit bounds for a geodesic observer were derived in Ref.~\cite{ko:15}. A finite bound is violated by the $1/f^2$ divergence of the energy density that results in the divergence of its smeared time average \cite{t:19,t:20}. Thus we are faced with the following conundrum: either accretion to a UCO can only occur before the first marginally trapped surface appears, and PBHs, once formed, can only evaporate, or semiclassical physics breaks down at the horizon scale. We restrict our discussion to evaporating PBHs in what follows.

Transformation to the orthonormal basis shows that depending on the behavior of the regular terms the EMT of these solutions belongs to type II or type III in the Segre\textendash{}Hawking\textendash{}Ellis classification scheme \cite{he:book, mmv:17}.  At $r\sim r_\sg$, the EMT coincides with that of a perfect exotic (i.e.\ NEC-violating) null fluid only if the metric is sufficiently close to Vaidya metrics (see Appendix \ref{aA}). However, this fluid is the key ingredient of matter near the apparent horizon and becomes dominant as $r\to  r_\sg$ for all  for all $\tau_a\sim f^{\,0}$ solutions.

Combining Eqs.~\eqref{intf} and \eqref{dv=0} allows to obtain an explicit form of the transformation between $(t,r)$ and $(v,r)$ coordinates near $r_\sg$ of an evaporating PBH. Since $t(v,r_++\delta y)=t(r_\sg)-\delta y/|r_\sg'| + \cdots$, we have
\be
	x(v,r_++y)=  -r''_\sg y^2/(2 r_\sg'{}^2)+\cO( y^3).   \label{xyrel}
\ee

\subsection{Extreme solution. Static solutions}
The static solution with $k=0$ is impossible, as in this case $\mathfrak{T}$ would diverge at the apparent horizon. Consequently, EMT components that allow for static solutions must behave differently. Many models of static nonsingular black holes assume  finite values of energy density and pressure at the horizon \cite{f:14,hay:06,coy:15}. With respect to the invariants of Eq.~\eqref{twoscal}, this is the $k=1$ solution, with
\be
\tau_t\to E(t)f, \qquad \tau^r\to P(t)f, \qquad \tau^r_{t}\to\Phi(t) f,   \label{rhop}
\ee
where $\rho=E$ and $p=P$ at the apparent horizon. Any two functions can be expressed algebraically in terms of the third and $8\pi r_\sg ^2E \leqslant 1$ to ensure that $C(t,r)-r_\sg>0$ for $r>r_\sg$. Appendix \ref{bA} provides a brief summary of their properties and gives explicit expressions for the metric functions $C\approx r_\sg+8\pi r_\sg^2Ex$ and $h$.

We now show that only a unique dynamic case with the extreme value of $E$ is possible. From Eqs.~\eqref{rhop}, \eqref{thev} and \eqref{cv1} it follows that $w_1=1$. As a result, as $\Delta_v(r)\defeq C_+(v,r)-r$ changes sign at $r=r_+$, the leading
terms in the expansion of the MS mass in Eq.~\eqref{cv1} are $C_+ = r_+ + y + w_3 y^3$, where $w_3 \leq 0$ and $y \defeq r-r_+$. If $w_3=0$ the nonlinear terms begin from a higher odd power.

This expression for the MS mass must coincide with $C\big(t(v,r_++y),r_++y)$. Eq.~\eqref{xyrel} also holds for $k=1$, hence  \be
C_+= C=r_\sg+y+(1-8\pi r_\sg^2 E)\frac{r''_\sg y^2}{2r'_\sg{}^2}+\cO(y^3),
 \ee
and thus $E\equiv 1/(8\pi r^2_\sg)$. Using the next terms in the expansion of $\tau_t$ leads to $f\approx c_{32}(t)x^{3/2}/r_\sg$ for some coefficient $c_{32}(t)>0$, setting via Eq.~\eqref{rhop} the scaling of other leading terms in the EMT. Consistency of Eqs.~\eqref{gtr} and \eqref{grr} implies $P=-E=-1/(8\pi r^2_\sg)$ and $\Phi=0$. From the next order expansion we obtain $h=-\tfrac{3}{2}\ln (x/\xi)+\cO(\sqrt{x})$ and the relation $r'_\sg=-c_{32}\xi^{3/2}/r_\sg$. Details of the calculation are presented in Appendix \ref{bA}.

On the other hand, solutions with a time-independent apparent horizon or general static solutions do not require $w_1=1$ to satisfy Eqs.~\eqref{thev}--\eqref{ther}. Since $r_+(v)=r_\sg(t)=\mathrm{const}$ it is possible to have non-extreme solutions. Then Eq.~\eqref{gtr} implies $\Phi=0$ and the identity $E=-P$ follows from Eq.~\eqref{ther}, leading to a regular function $h(t,r)$. However, in this case Eq.~\eqref{intf} indicates that the apparent horizon cannot be reached in a finite time $t$.

\section{Physical black holes in modified gravity}
There are numerous arguments as to why a classical theory of gravity may or should differ from GR \cite{burgess:04}. Strong fields in the vicinity of UCOs are one of the regimes where the effects of modified gravity are expected to be discernable. Mathematically, these theories are typically more involved than GR, and exact and approximate black hole solutions are used both to test the consistency of such theories and also to differentiate between models of horizonless UCOs and PBHs \cite{ckms:18}.

One group of models includes various additional curvature-dependent terms in the gravitational Lagrangian, $\eL_\mathrm{g}=R+\lambda \eF(g^{\mu\nu}, R_{\mu\nu\rho\sigma})$, where $\lambda$ is a small dimensionless parameter
\cite{burgess:04,cdl:11,f-R}. The Einstein equations are modified by fourth or higher-order terms, $G_{\mu\nu} + \lambda \EuScript{E}_{\mu\nu} = 8 \pi T_{\mu\nu}$, where the terms $\EuScript{E}_{\mu\nu}$ result from the variation of $\eF$ \cite{pad:11}. The most general spherically symmetric metric is still given by Eq.~\eqref{sgenm}, and the requirements of finiteness of $\mathrm{T}$ and $\mathfrak{T}$ are still meaningful. However, they are no longer directly related to the finiteness of the curvature scalars. For example, in $\mathfrak{f}(R)$ theories \cite{f-R}, $\eL=\mathfrak{f}(R)$,
\be
	\maf'(R)R+2\maf(R)+3\square\maf'(R)=8\pi\mathrm{T},
\ee
and unlike in GR the finiteness of $T^\theta_{\,\theta}$ is not guaranteed a priori. It is conceivable that the metric is such that the curvature invariants are finite, but $\square R$ and thus $\mathrm{T}$ diverge at the apparent horizon.

Nevertheless, the two types of solutions discussed above are the only perturbatively possible classes of solutions in spherical symmetry. While their existence must be established separately for each theory, it is clear that divergences stronger than those allowed in GR are not permitted at any order of $T_{\mu\nu}=\bar{T}_{\mu\nu}+\lambda T^{(1)}_{\mu\nu}+\cdots$, where $\bar{T}_{\mu\nu}$ denotes the unperturbed GR expression. Such terms will contribute stronger singularities to the functions $C$ and $h$, and thus invalidate the perturbative expansion close to the apparent horizon.

The Einstein\textendash{}Cartan theory of gravity is a modification of GR in which spacetime can have torsion in addition to curvature \cite{cdl:11,ect}. The torsion tensor is expressed as the antisymmetric part of the connection $Q^\mu_{\nu\eta}=\half(\Gamma^\mu_{\nu\eta}-\Gamma^\mu_{\eta\nu})$. Despite having a non-metric part of the connection, it is still assumed that $\nabla\sg_{\mu\nu}=0$. The full set of equations now consists of the equations for $G_{\mu\nu}$ that are related to the EMT, and the equations for $Q^\mu_{\nu\eta}$ that relate the torsion to the density of intrinsic angular momentum.

However, it is possible to represent this system by a single set of Einstein equations with an effective EMT on the rhs,
\be
	\mathring{G}_{\mu\nu}=8\pi T^\mathrm{eff}_{\mu\nu},
\ee
where $\mathring{G}_{\mu\nu}$ is derived from the metric alone and the effective EMT includes terms that are quadratic in spin \cite{ect,tr:06}. Requiring now that $\mathring R^\mu_{\,\nu}$  and $\mathring R^{\mu\nu}\mathring R_{\mu\nu}$ are finite at the apparent horizon $r=r_\sg$ leads to the same types of PBH solutions.

\section{Implications}
\subsection{Black hole formation}
Consider now possibilities for horizon formation. Assume that the first marginally trapped surface appears at some $v_\mathrm{S}$ at $r=r_+(\vS)$. For $v\leqslant \vS$ the MS mass in its vicinity can be described by modifying Eq.~\eqref{cv1} as
\be
	C(v,r)=\sigma(v)+r_*(v)+\sum_{i\geqslant 1} w_i(v)(r-r_*)^i,
\ee
where the deficit function $\sigma(v)\defeq C(v,r_*)-r_*\leqslant 0$, and $r_*(v)$ corresponds to the maximum of $\Delta_v(r)\defeq C(v,r)-r$. At the advanced time $\vS$ the location of the maximum corresponds to the first marginally trapped surface, $r_*(\vS)=r_+(\vS)$ and $\sigma(\vS)=0$. For $v\geqslant\vS$ the MS mass is described by Eq.~\eqref{cv1}. For $v\leqslant \vS$ the (local) maximum of  $\Delta_v$   satisfies $d\Delta_v/dr=0$, hence $w_1(v)-1\equiv 0$. Before the PBH is formed there are no a priori restrictions on the evolution of $r_*$. However, since an accreting PBH leads to a firewall, $r'_+(\vS)\leqslant 0$. Since the trapped region is of a finite size for $v>\vS$ the maximum of $C(v,r)$ does not coincide with $r_+(v)$. As a result, $w_1(v)<1$ for $v>\vS$.

This scenario means that at its formation a PBH is described by a $k=1$ solution that is necessarily extreme. This is due to Eq.~\eqref{thev}, as 
\be
	\theta_v(v,r_+)=(1-w_1)\frac{r'_+}{8\pi r^2_+}.
\ee
It immediately switches to the $k=0$ solution. Since the energy density and pressure are negative in the vicinity of the apparent horizon and positive in the vicinity of the inner horizon \cite{f:14,t:19}, density and pressure jump at the intersections of the two horizons. However, an abrupt transition from $f^1$ to $f^0$ behavior is only of conceptual importance: this aspect of the evolution is continuous in $(v,r)$ coordinates and there will be no discontinuity according to observers crossing the $r=r_*$ and subsequently $r=r_\sg$ surfaces. We discuss the relevant timescales in Sec.~\ref{disc}.

\subsection{Surface gravity}    \label{surfg}
The surface gravity $\kappa$ plays an important role in GR and semiclassical gravity \cite{he:book,fn:book,faraoni:b}. However,
it is unambiguously defined only in stationary spacetimes, where it is related to the properties of a Killing horizon (for the
Schwarzschild black hole $\kappa=(2r_\sg)^{-1}$), and is proportional to the Hawking temperature.

Emission of Hawking-like radiation does not require the formation of an event or even an apparent horizon \cite{pp:09,haj:87,blsv:06,vsk:07}. This pre-Hawking radiation is related to the peeling property of null geodesics \cite{peel:blsv}. The peeling surface gravity $\kappa_\mathrm{peel}$  is defined from the near-horizon behavior of the null geodesics \cite{faraoni:b,kappa} as the linear coefficient in the Taylor expansion of the equation $ {dr}/{dt}=\pm e^{h}f$. However, such an expansion is impossible for the metric functions of Eqs.~\eqref{cv1}\textendash{}\eqref{hv1}. Alternatively, to be compatible with Eq.~\eqref{dv=0}, $\kappa_\mathrm{peel}$ should be infinite, as we now demonstrate.

Consider an observer Eve at some fixed areal radius $r$. Her four-velocity is $u_\rE^\mu = \delta^\mu_0 / \sqrt{-\sg_{00}}$ and her four-acceleration $a_\rE^\mu=(0,\Gamma^r_{tt}/\sg_{00},0,0)$ satisfies
\be
	g^2\defeq a^\mu_{\rE} a_{{\rE}\mu}=\frac{|r_\sg'|}{16\sqrt{\xi}x^{3/2}} + \cO(x^{-1}),
	\label{g2tr}
\ee
where $x=r-r_\sg$ and we have used Eqs.~\eqref{k0met} and \eqref{rpr1}. The same dependence is obtained if the equivalent $(v,r)$ metric is used. A naive attempt to generalize the surface gravity by using the expression
\be
	\kappa_z=\lim_{r\to r_\sg}zg,
\ee
where $z=-\sqrt{\sg_{00}}$ is the redshift factor that agrees with the standard definition in static spacetimes, diverges as
\be
	zg=\frac{|r'_\sg|}{4x}+\cO(x^{-1/2}).
\ee

The peeling surface gravity $\kappa_\mathrm{peel}$  is the coefficient of the linear term in the expansion of
\be
	\frac{dr}{dt}=\pm e^{h}f(t,r)
\ee
in powers of $x$ as $x\to 0$ (i.e.\ $r\to r_\sg$). For differentiable $C$ and $h$ the result is
\be
	\kappa_\mathrm{peel} = \frac{e^{h(t,r_\sg)} \left( 1 - C'(t,r_\sg) \right)}{2 r_\sg} .
\ee
However, for $k=0$ solutions this quantity is undefined as for a radial geodesic Eq.~\eqref{dv=0} implies \be
	\frac{dr}{dt} = \pm|r_\sg'| + \cO(\sqrt{x}),
\ee
providing a constant term that is absent from the derivation of  $\kappa_\mathrm{peel}$ above.

\section{Discussion}  \label{disc}
The remarkable properties of a PBH follow solely from the regularity of its apparent horizon and finite formation time according to Bob. The NEC is violated in the vicinity of the apparent horizon and the matter content is dominated by a null fluid. If the semiclassical picture is valid, then accretion leads to a firewall that violates the bounds on the violation of the NEC. As a result, accretion can occur only before a PBH is formed. This firewall is not an artifact of spherical symmetry: the same effect was demonstrated for the Kerr\textendash{}Vaidya metric \cite{kdt:20}. While a sufficiently slow massive test particle can be prevented from crossing the horizon, the crossing generally happens in finite time according to a distant observer. Taking the proper radial velocity to be of the order of one (for a test particle falling from infinity with zero initial velocity into a Schwarzschild black hole $\dot r (\tau)=-3/4$ at $r=r_\sg$), we see that the time dilation for a nonstationary Alice is $dt_\mathrm{A}/d\tau\sim |r'_\sg|^{-1}$ at the apparent horizon.

On the other hand, it is still not clear how the collapsing matter actually behaves. Violation of the NEC requires some mechanism that converts the original matter into the exotic matter present in the vicinity of the forming apparent horizon, thereby creating something akin to a shock wave to restore the normal behavior near the inner horizon. However, emission of the collapse-induced radiation \cite{pp:09,haj:87,blsv:06,vsk:07,peel:blsv} is a nonviolent process that approaches at latter times the standard Hawking radiation and Page's evaporation law $r_\sg'=-\alpha/r_\sg^2$, $\alpha \sim 10^{-3}-10^{-4}$ \citep{fn:book,dnP.1976}. Moreover, divergence of $\kappa_\mathrm{peel}$ requires a careful reevaluation of Hawking-like radiation by PBHs. Hence the observed UCOs may actually be horizonless --- not due to some exotic supporting matter or dramatic variation in the laws of gravity, but simply because the conditions for the formation of a PBH have not been met at the present moment of $t$.

Even if the necessary NEC violation occurs in nature, the process may be too slow to transform the UCOs that we observe into PBHs. Eq.~\eqref{dv=0} sets the timescale of the last stages of infall according to Bob. Assuming that it is applicable through the radial interval of the order of $r_\sg$, we have $t_\mathrm{in}\sim r_\sg/r'_\sg$. For an evaporating macroscopic PBH, this is of the same order of magnitude as the Hawking process decay time $t_\mathrm{evp}\sim 10^3 r_\sg^3$. Such behavior was found in thin shell collapse models, where the exterior geometry is modeled by a pure outgoing Vaidya metric \cite{bmt:19}. For a solar mass black hole this time is about $10^{64}$ yr, indicating that it is simply too early for the horizon to form. It is also conceivable that the conditions are not met before evaporation is complete or before effects of quantum gravity become dominant \cite{fn:book, coy:15}.

The possibility that exotic new physics is only needed for the formation of black holes, but not for the formation of horizonless objects, has interesting consequences for the information loss paradox \cite{fn:book,rev-0,gna:17}. Its formulation is ineluctably linked to the existence of an event horizon and singularity \cite{visser}. However, horizon avoidance \cite{bht:17} --- and thus elimination of the paradox --- may occur due to the absence of new physics, and not because of it. A better understanding of the near-horizon geometry of PBHs will improve models developed to take full advantage of the new era of multimessenger astronomy \cite{bh-map,gna:17}, using observations not only to learn about the true nature of astrophysical black holes, but also to obtain new insights into fundamental physics.

\acknowledgments
We thank Valetina Baccetti, Gary Gibbons, Viqar Hussain, Eleni Kontou and Robert Mann for useful discussions and helpful comments. SM is supported by an International Macquarie University Research Excellence Scholarship and a Sydney Quantum Academy Scholarship. The work of DRT was supported in part by the Southern University of Science and Technology, Shenzhen, China, and by the ARC Discovery project grant DP210101279.

\appendix
\section{Some properties of $\boldsymbol{k=0}$ solutions}\label{aA}
Using Eqs.~\eqref{cv1} and \eqref{hv1} the EMT components at the apparent horizon $r=r_+(v)$ are
\be
	\theta_v=\frac{(1-w_1)r_+'}{8\pi r_+^2}, \quad \theta_{vr}=-\frac{w_1}{8\pi r_+^2}, \quad  \theta_r=\frac{\chi_1}{4\pi r_+}.
\ee
The EMT expansion for $k=0$ solutions in $(t,r)$ coordinates is given by
\begin{align}
\tau_t&=-\Upsilon^2 +\sum_{j \geqslant 1} \alpha_{j/2}x^{j/2}, \\
\tau_t^r&=-\Upsilon^2 +\sum_{j \geqslant 1}\beta_{j/2}x^{j/2},\\
\tau^r&=-\Upsilon^2 +\sum_{j \geqslant 1}\gamma_{j/2}x^{j/2}.
\end{align}
The limit $\tau_t\to \tau^r\to-\Upsilon^2$, i.e.\ violation of the NEC in the vicinity of $r_\sg(t)$, is required to ensure that a real spherically symmetric solution with apparent horizon exists, i.e.\ that the equation $C(t,r)=r$ has a solution for finite $t$.

This result should be compared with the conclusions of Sec.\ 9.2 of Hawking and Ellis \cite{he:book}, where it was shown that in general asymptotically flat spacetimes with an asymptotically predictable future the trapped surface cannot be visible from future null infinity unless the weak energy condition is violated. Our result \cite{bmmt:19} is on the one hand more restrictive as its derivation uses spherical symmetry, but on other hand does not require any assumptions about the asymptotic structure of spacetime. The triple limit $\tau_a \to-\Upsilon^2$ was observed in ab initio calculations of the renormalized EMT on a
Schwarzschild background \cite{leviori:16}.

The first constraint can be read off from Eq.~\eqref{ther}, giving
\be
\alpha_{12}+\gamma_{12}-2\beta_{12}=0,   \label{transcon}
\ee
where we have omitted the  slash symbol from fractional indices to reduce clutter. Additional constraints follow from the higher-order expansion of Eq.~\eqref{gtr}.

The higher-order terms in the expansion
\be
	t(v,r_++y)=t(v,r_+)+y/r_\sg'+\cO(y^2)
\ee
are obtained by taking the limit $r\to r_\sg$ of the corresponding derivatives of $dt/dr=-(e^h f)^{-1}$. As a result
\be
	x=0\to x=y-r_\sg(v,r_++y) = - \frac{r_\sg'' y^2}{2 r_\sg'^2} + \cO(y^3).
\ee
Invariance of the MS mass, $C_+=C$, then allows to identify the terms via
\be
	r_++w_1y+\cdots =r_\sg +\left(1-\frac{r_\sg\sqrt{-r_\sg''}}{\sqrt{2\xi}}\right) y+\cdots,    \label{msexpan}
\ee
where we have used Eqs.~\eqref{rpr1} and \eqref{xyrel} to simplify the coefficient of $y$ on the rhs.

Curvature scalars can be conveniently evaluated using the expression for the Riemann tensor in the orthonormal frame that is based on the normalization $(\pad_t,\pad_r,\pad_\theta,\pad_\varphi)$ \cite{av:10}. In particular, the Kretschmann scalar $K\defeq R_{\mu\nu\eta\zeta}R^{\mu\nu\eta\zeta}$ satisfies the simple expression
\be
	K = 4 R_{\hat{0}\hat{1}\hat{0}\hat{1}}^2+8R_{\hat{0}\hat{2}\hat{0}\hat{2}}^2-16R_{\hat{0}\hat{2}\hat{1}\hat{2}}^2+8R_{\hat{1}\hat{2}\hat{1}\hat{2}}^2+4R_{\hat{2}\hat{3}\hat{2}\hat{3}}^2 .
\ee

The EMT in this orthonormal basis has the form
\be
	T_{\hat{a}\hat{b}}=\left(\begin{tabular}{cc|cc}
 $q+\mu_1$ & $q+\mu_2$  & $0$& $0$\\
$q+\mu_2$ & $q+\mu_3$  & $0$& $0$ \\ \hline \label{IVP}
$0$ & $0$ & $\EuScript{P}$ & 0 \\
$0$ & $0$ & 0 & $\EuScript{P}$
 \end{tabular}\right),
\ee
where
\be
	q=-\frac{\Upsilon}{4\sqrt{\pi r_\sg x}},
\ee
and the remaining coefficients are finite at the apparent horizon. Analogous expressions can be obtained using $(v,r)$ coordinates.

For an evaporating black hole, if the geometry is approximately Vaidya ($w_1=0$, $\chi_1=0$), then
\be
	\rho_{\mathrm{A}}^<=p_{\mathrm{A}}^<=\phi_{\mathrm{A}}^<=-\frac{\Upsilon^2}{4 \dot{r}_A^2},\label{comov}
\ee
at $r_\mathrm{A}=r_\sg=r_+$ \cite{t:19}, where $\dot{r}_\mathrm{A} \equiv d r_\mathrm{A} / d\tau$.

For the MS mass $C(u)$, where $u$ is the retarded null coordinate, solutions with $C'(u)<0$ satisfy the NEC and only a complex-valued transformation to $(t,r)$ coordinates exists. The same applies to $C(v)$, where $v$ is the advanced null coordinate, if $C'(v)>0$ \cite{bmmt:19}.

\section{Some properties of the fully regular $\boldsymbol{k=1}$ solution} \label{bA}
For nonextreme solutions the EMT expansion is
\begin{align}
\tau_t &= E f + \sum_{j\geqslant 4}\alpha_{j/2}x^{j/2}, \label{eb1}\\
\tau_t^r &= \Phi f + \sum_{j\geqslant 4}\beta_{j/2}x^{j/2},\label{eb2}\\
\tau^r &= P f + \sum_{j\geqslant 4}\gamma_{j/2}x^{j/2}.\label{eb3}
\end{align} 
The leading terms of the metric functions are
\begin{align}
C&=r_\sg+8\pi r_\sg^2 E x + \cO(x^{3/2}),\\
 h&=-\ln\frac{x}{\xi}+\cO(\sqrt{x}),
\end{align}
where $8 \pi r_\sg^2 E \leqslant 1$ due to the definition $C(t,r_\sg)=r_\sg$, $f>0$ for $r>r_\sg$. The functions $P$ and $\Phi$ can be expressed as
\be
	P=\frac{-1+4\pi r_\sg^2 E}{4\pi r_\sg^2}, \qquad \Phi=\pm\frac{1-8\pi r_\sg^2 E}{8\pi r_\sg^2}.
\ee

However, only the extreme case $E=1/(8\pi r_\sg^2)$ is consistent for an evolving apparent horizon. Since the limit of Eq.~\eqref{thevr} results in
\be
	-\frac{w_1}{8\pi r_+^2}=\Phi-E,
\ee
we have $\Phi=0$ and by Eq.~\eqref{ther} $P=-E=- 1/(8\pi r_\sg^2)$. As a result $f\approx c_{32}x^{3/2}/r_\sg$ near $r=r_\sg$, and the next term in the expansion of $\tau_t$ is $\alpha_2$. Consistence with Eq.~\eqref{thevr} imposes $\beta_2=\alpha_2$.

Solving Eq.~\eqref{gtt} results in
\be
	C(t,r)=r-4\sqrt{-\pi \alpha_2/3}\,r_\sg^{3/2} x^{3/2}+\cO(x^2),  \label{cextr}
\ee
i.e.\ $\pad_t C= \tfrac{3}{2}r_\sg'c_{32}\sqrt{x}$ , and 
\be
	\alpha_2=-\frac{3 c_{32}^2}{16 \pi r_\sg^3}.    \label{al2}
\ee

On the other hand, the leading term of the flux $\tau^r_t$ is $\beta_2 x^2$, and to satisfy Eq.~\eqref{gtr} the function $h$ that results from Eq.~\eqref{grr} should satisfy
\be
	\pad_x h=-\frac{3}{2x} =4\pi\frac{(\alpha_2+\gamma_2)r_\sg^3}{c_{32}^2 x}
\ee
at leading order. Using Eq.~\eqref{al2} we have $\gamma_2=\alpha_2$, and according to Eq.~\eqref{ther} $\beta_2=\alpha_2$. At leading order
\be
	h=-\frac{3}{2}\ln\frac{x}{\xi}+\cO(\sqrt{x}),
\ee
and Eq.~\eqref{gtr} leads to
\be
	r'_\sg=- c_{32}\xi^{3/2}/r_\sg,
\ee
as we consider only evaporation. Direct evaluation of $R$, $R_{\mu\nu}R^{\mu\nu}$ and $K$  shows that this condition suffices to ensure their finite values on the apparent horizon.

\end{document}